\documentstyle[12pt]{article}
\def\be{\begin{equation}}
\def\ee{\end{equation}}
\def\bea{\begin{eqnarray}}
\def\eea{\end{eqnarray}}

\def\ra{\rightarrow}

\def\bar{\overline}

\def\a{\alpha}
\def\b{\beta}

\def\e{\epsilon}

\def\l{\lambda}

\def\bc{\begin{center}}
\def\ec{\end{center}}

\def\O{{\cal O}}
\def\PR#1#2#3{Phys. Rev.  {\bf #1}, (#3) #2}
\def\PRL#1#2#3{Phys. Rev. Lett. {\bf #1}, (#3) #2}
\def\PL#1#2#3{Phys. Lett. {\bf #1}, (#3) #2}

\def\NP#1#2#3{Nucl. Phys. {\bf #1}, (#3) #2}

\def\PTP#1#2#3{Prog. Theor. Phys. {\bf #1}, (#3) #2}

\begin{document}
\title{Large Lepton Flavor Mixings \\
in $SU(6)\times SU(2)_R$ Model}
\author{
{M. Matsuda$^1$}\thanks{E-mail address:mmatsuda@auecc.aichi-edu.ac.jp}
{ and T. Matsuoka$^2$}\thanks{E-mail
address:matsuoka@kogakkan-u.ac.jp}
\\
\\
\\
{\small \it $^1$Department of Physics and Astronomy,
Aichi University of Education,}\\
{\small \it Kariya, Aichi, 448-8542 Japan}\\
{\small \it $^2$Kogakkan University, Nabari, Mie, 518-0498 Japan}}
\date{}
\maketitle
\vspace{-10.5cm}
\begin{flushright}
hep-ph/0009077 \phantom{M}
AUE-00-03 \phantom{M}
KGKU-00-03 
\end{flushright}
\vspace{10.5cm}
\vspace{-2.5cm}
\begin{abstract}
The lepton masses and mixings are studied on the basis of 
string inspired $SU(6)\times SU(2)_R$ model with global 
flavor symmetries. 
Provided that sizable mixings between lepton doublets $L$ and 
Higgsino-like fields $H_d$ with even $R$-parity occur and 
that seesaw mechanism is at work in the neutrino sector, 
the model can yield a large mixing angle solution with 
$\tan \theta_{12},\ \tan \theta_{23} = \O(\sqrt{\l})$ 
$(\l \simeq 0.22)$, which is consistent with the recent experimental 
data on atmospheric and solar neutrinos. 
In the solution Dirac mass hierarchies in the neutrino sector 
cancel out with the heavy Majorana sector in large part 
due to seesaw mechanism. 
Hierarchical pattern of charged lepton masses can be also 
explained. 
\end{abstract}

{\sf keyword:MNS matrix, large mixing, unification model}
\newpage

Recent experimental data have suggested large lepton flavor mixings. 
In fact, the latest atmospheric neutrino results from Super-Kamiokande 
are consistent with $\nu_{\mu}$-$\nu_{\tau}$ oscillation with 
$\sin^2 2\theta_{23} > 0.88$ and $\Delta m^2_{23} = (1.5 \sim 5) \times 
10^{-3} {\rm eV}^2$\cite{Atmos}. 
The latest solar neutrino results from Super-Kamiokande are in favor of 
the large mixing angle MSW region for $\nu_e$-$\nu_{\mu}$ oscillation 
with $\sin^2 2\theta_{12} \sim 0.75$ and $\Delta m^2_{12} \sim 2.2 \times 
10^{-5} {\rm eV}^2$\cite{Solar}. 
These results indicate that lepton flavor mixing matrix (MNS matrix) 
is remarkably different from quark flavor mixing matrix (CKM matrix) 
in their hierarchical structure. 
At first sight it seems that the distinct flavor mixings of quarks 
and leptons are in disaccord with the quark-lepton unification. 
However, in a wide class of unification models, 
the situation is not so simple. 
This is because the massless sector in the supersymmetric unification 
theory includes extra particles beyond the standard model 
and then there may occur extra-particle mixings such as 
between quarks (leptons) and colored Higgsino-like 
fields (doublet Higgsino-like fields) with even $R$-parity. 
In order to study fermion masses and mixings we have to take 
into account the effects of such extra-particle mixings. 
In addition, in the neutrino sector we should incorporate 
the extra-particle mixings with seesaw mechanism
\cite{Seesaw}. 
In the previous paper\cite{CKM} we have shown that 
the observed hierarchical structure of quark masses and CKM matrix 
can be naturally understood in $SU(6) \times SU(2)_R$ model. 
In short, after integrating out heavy modes which get masses 
at intermediate energy scales, 
we have found that the extra-particle mixings possibly 
cause Yukawa hierarchies to change significantly. 
In this paper we study characteristic features of lepton masses and 
MNS matrix in the context of the string 
inspired $SU(6) \times SU(2)_R$ model with global flavor symmetries.

\vspace{1cm}

The model discussed here is the same as in 
Ref.\cite{CKM,Matsu1,Matsu2,Matsu3}. 
Here we enumerate main points of the present model. 
\begin{enumerate}
\item We choose $SU(6) \times SU(2)_R$ 
as the unification gauge symmetry at the string scale $M_S$, 
which can be derived from the perturbative heterotic 
superstring theory via the flux breaking\cite{Matsu4}. 

\item Matter superfields consist of three family and one 
vector-like multiplet, i.e., 
\be
  3 \times {\bf 27}(\Phi_{1,2,3}) + 
        ({\bf 27}(\Phi_0)+\overline{\bf 27}({\bar \Phi})) 
\ee
in terms of $E_6$. 
Under $G= SU(6) \times SU(2)_R$, the superfields $\Phi$ in 
{\bf 27} of $E_6$ are decomposed into two groups as 
\be
  \Phi({\bf 27})=\left\{
       \begin{array}{lll}
         \phi({\bf 15},{\bf 1})& : 
               & \quad \mbox{$Q,L,g,g^c,S$}, \\
          \psi(\overline{\bf 6},{\bf 2}) & : 
               & \quad \mbox{$(U^c,D^c),(N^c,E^c),(H_u,H_d)$}, 
       \end{array}
       \right.
\label{27}
\ee
where $g$, $g^c$ and $H_u$, $H_d$ represent colored Higgs and 
doublet Higgs fields, respectively. 
$N^c$ is the right-handed neutrino superfield and 
$S$ is an $SO(10)$-singlet. 
It is noticeable that under $G$ doublet Higgs and color-triplet 
Higgs fields belong to  different representations. 
This situation is favorable to solve the triplet-doublet 
splitting problem. 
Although $D^c$ and $g^c$($L$ and $H_d$) 
have the same quantum numbers under the standard model 
gauge group $G_{SM} = SU(3)_c \times SU(2)_L \times U(1)_Y$, 
they belong to different irreducible representations of $G$. 

\item We assign odd $R$-parity for $\Phi_{1,2,3}$ and even for $\Phi_0$ 
and $\overline{\Phi}$, respectively. 
Since ordinary Higgs doublets have even $R$-parity, 
they belong to $\Phi_0$. 
It is assumed that $R$-parity remains unbroken down to 
the electroweak scale. 

\item The gauge symmetry $G$ is assumed to be spontaneously broken 
at $|\langle \phi_0({\bf 15, \ 1})\rangle |$ and 
subsequently at 
$|\langle \psi_0({\bf \overline{6}, \ 2})\rangle |$. 
Since the fields which develop non-zero VEV's are singlets under 
the remaining gauge symmetries, 
they are assigned as 
$\langle \phi_0({\bf 15, \ 1})\rangle = \langle S_0 \rangle $ and 
$\langle \psi_0({\bf \overline{6}, \ 2})\rangle = \langle N^c_0 \rangle $.
The $D$-flat conditions require 
$\langle S_0\rangle=\langle {\overline S} \rangle$ 
and $\langle N_0^c\rangle=\langle {\overline N^c} \rangle$ at each step 
of the symmetry breakings 
\be
   G = SU(6) \times SU(2)_R 
     \buildrel \langle S_0 \rangle \over \longrightarrow 
             SU(4)_{\rm PS} \times SU(2)_L \times SU(2)_R  
     \buildrel \langle N^c_0 \rangle \over \longrightarrow 
     G_{SM}, 
\ee
where $SU(4)_{\rm PS}$ represents the Pati-Salam 
$SU(4)$\cite{Pati}. 
Hereafter it is supposed that the symmetry breaking scales 
are roughly $\langle S_0 \rangle = 10^{17 \sim 18}$GeV and 
$\langle N^c_0 \rangle = 10^{15 \sim 17}$GeV. 
In the present model the symmetry breakings at such large scales 
can be realized\cite{Scale}. 
At the first step of the symmetry breaking fields $Q_0$, $L_0$, 
${\overline Q}$, ${\overline L}$ and $(S_0 - {\overline S})/\sqrt{2}$ 
are absorbed by gauge fields. 
Through the subsequent symmetry breaking fields $U_0^c$, $E_0^c$, 
${\overline U}^c$, ${\overline E}^c$ and 
$(N_0^c - {\overline N}^c)/\sqrt{2}$ are absorbed. 

\item Gauge invariant trilinear couplings in the superpotential 
$W$ become to be of the forms 
\bea
    (\phi ({\bf 15},{\bf 1}))^3 & = & QQg + Qg^cL + g^cgS, 
\label{eqn:W1}
                                                            \\
    \phi ({\bf 15},{\bf 1})(\psi (\overline{\bf 6},{\bf 2}))^2 & 
            = & QH_dD^c + QH_uU^c + LH_dE^c  + LH_uN^c 
                                            \nonumber \\ 
             {}& & \qquad   + SH_uH_d + 
                     gN^cD^c + gE^cU^c + g^cU^cD^c.
\label{eqn:W2}
\eea 

\item We introduce a global flavor symmetry ${\bf Z_N}$. 
Then the Froggatt-Nielsen mechanism is at work for the 
interactions\cite{F-N}. 
The string theory naturally provides the discrete symmetry 
stemming from the symmetric structure of the compactified space. 
The stringy discrete symmetry may be either $R$ symmetry or 
non-$R$ symmetry. 
Here we take a non-$R$ discrete symmetry. 
\end{enumerate}

In the present framework the effective Yukawa interactions for 
charged leptons are of the form\cite{Matsu1} 
\be
  M_{ij} \, L_i E^c_j H_{d0} 
\label{QUH}
\ee
with 
\be
    M_{ij} = (M_0)_{ij} \left( \frac{\langle X \rangle}
                              {M_S}\right)^{e_{ij}} 
           = m_{ij} \, x^{e_{ij}}, 
\label{Mij}
\ee
where the subscripts $i$ and $j$ stand for the generation 
indices and the coupling constants $m_{ij}$'s are assumed 
to be ${\cal O}(1)$ with rank\,$m_{ij}=3$. 
$X \equiv (S_0{\bar S})/M_S$ is a singlet with a nonzero flavor 
charge and $x \equiv \langle X \rangle /M_S < 1$. 
The exponents $e_{ij}$ are some non-negative integers which 
are settled by the flavor symmetry. 
Yukawa hierarchies are derived by assigning appropriate 
flavor charges to the matter fields. 
Concretely, when $a_i$, $b_i$ $(i=0,1,2,3)$ and ${\bar a}$, ${\bar b}$ 
denote the flavor charges of matter fields $\phi _i$, $\psi _i$ 
and ${\bar \phi }$, ${\bar \psi }$, respectively, 
the singlet $X$ has its charge $a_X = a_0 + {\bar a}$. 
Provided that $|a_X|$ and $N$ are prime with each other, 
we can take $a_X = -1$ without loss of generality. 
In this case the flavor symmetry yields the relation 
\be
   e_{ij} = a_i + b_j + b_0 \qquad {\rm mod}\ \  N 
\ee
for the above effective Yukawa interactions. 
Hereafter we use the notation $\a_i$ and $\b_i$ $(i=1,2,3)$ defined by 
\be
  \a_i = a_i - a_3, \qquad  \b_i = b_i - b_3. 
\ee
By definition we have $\a_3 = \b_3 = 0$. 
Assuming 
\be
   e_{33} = a_3 + b_3 + b_0 = 0 \qquad {\rm mod} \ \ N, 
\ee
we have a $3 \times 3$ mass matrix 
\be
M = \left(
  \begin{array}{ccc}
 m_{11}x^{\a_1+\b_1} & m_{12}x^{\a_1+\b_2} & m_{13}x^{\a_1} \\
 m_{21}x^{\a_2+\b_1} & m_{22}x^{\a_2+\b_2} & m_{23}x^{\a_2} \\
 m_{31}x^{\b_1}      & m_{32}x^{\b_2}      & m_{33} 
  \end{array}
  \right).
\label{eqn:Mu}
\ee
By virtue of $SU(6) \times SU(2)_R$ gauge symmetry 
the up-type quark mass matrix is given by the same matrix $M$. 
The assumption $e_{33} = 0$ implies that 
the Yukawa coupling for top-quark is ${\cal O}(1)$. 
In the previous paper\cite{CKM} we showed that hierarchical 
pattern of quark masses and mixings can be reproduced by 
taking 
\be
  x^{\a_1} = \l^3, \qquad x^{\a_2} = \l^2, \qquad 
  x^{\b_1} = \l^4, \qquad x^{\b_2} = \l^2  
\ee
with $\l\simeq 0.22$. 
Hereafter we take this choice of the parameters. 

\vspace{1cm}

Below the scale $\langle N_0^c \rangle$ there appear both 
$L$-$H_d$ and $D^c$-$g^c$ mixings. 
Due to $L$-$H_d$ mixings the charged lepton mass matrix 
is expressed in terms of the $6 \times 6$ matrix 
\be 
\begin{array}{r@{}l} 
   \vphantom{\bigg(}   &  \begin{array}{ccc} 
          \quad   H_u^+   &  \quad  E^{c+}  &  
        \end{array}  \\ 
\widehat{M}_l = 
   \begin{array}{l} 
        H_d^-  \\  L^-  \\ 
   \end{array} 
     & 
\left( 
  \begin{array}{cc} 
       y_S H    &    0       \\
       y_N M    &  \rho _d M 
  \end{array} 
\right) 
\end{array} 
\label{eqn:Mlh} 
\ee 
in units of the string scale $M_S$. 
Three nonzero $3 \times 3$ matrices arise from 
the mass terms $H_{ij}\,{H_d}_i{H_u}_j\langle S_0 \rangle$, 
$M_{ij}\,L_i{H_u}_j\langle N_0^c \rangle$ and 
$M_{ij}\,L_iE_j^c\langle H_{d0} \rangle$, 
where 
\be
   H_{ij}=(H_0)_{ij}\, x^{\b_i + \b_j + \xi}
                          = h_{ij}\, x^{\b_i + \b_j + \xi}
\ee
with $ h_{ij}={\cal O}(1)$. 
The exponent $\xi$ represents the flavor charge of 
the trilinear products ${H_d}_3 {H_u}_3 S_0$, i.e., 
$\xi = 2b_3 + a_0$. 
Here $\xi$ is taken to be non-negative so that the trilinear 
couplings $H_{ij}\,{H_d}_i{H_u}_j S_0$ have the Yukawa hierarchy 
$|H_{1j}| \ll |H_{2j}| \ll |H_{3j}|$ for $j=1,\,2,\,3$ 
similar to $M_{ij}$. 
Here we use the notations $y_S$, $y_N$ and $\rho _d$ for 
the VEV's $\langle S_0 \rangle$, $\langle N^c_0 \rangle$ and 
$\langle H_{d0} \rangle = v_d$ in $M_S$ units, 
respectively. 
{}From Eq.(\ref{eqn:W2}) it is found that the matrix $H$ is 
symmetric. 
Since $\rho _d$ is very small compared to $y_S$ and $y_N$, 
the mixings between $E^c$ and $H_u$ are negligibly small. 
While the large mixings between $L$ and $H_d$ can 
occur depending on the relative magnitude of $y_S H$ and $y_N M$.

The matrix $\widehat{M}_l$ can be diagonalized by 
a bi-unitary transformation as 
\begin{equation} 
    \widehat{\cal V}_l^{-1} \widehat{M}_l \, 
                        \widehat{\cal U}_l. 
\label{eqn:Ml} 
\end{equation} 
To solve the eigenvalue problem, it is more instructive for us to 
take $\widehat{M}_l^{\dag} \widehat{M}_l$ expressed as 
\be
   \widehat{M}_l^{\dag} \widehat{M}_l =  \left(
  \begin{array}{cc}
     A_l + B_l   &   \e_d B_l   \\
     \e_d B_l    &  \e_d^2 B_l  
  \end{array}
  \right),
\ee
where $A_l = y_S^2 H^{\dag} H$ and $B_l = y_N^2 M^{\dag}M$ 
with $\e_d \equiv \rho_d / y_N$. 
Since $\e_d$ is a very small number, 
we can carry out our calculation by using perturbative 
$\e_d$-expansion. 
Among six eigenvalues three of them are given by eigenvalues of 
$(A_l + B_l)$ at the leading order, 
which represent heavy modes with the GUT scale masses. 
The remaining three are derived from diagonalization of $\e_d^2$-terms, 
i.e., 
\be
   \e_d^2 B_l - \e_d B_l \frac{1}{A_l + B_l} \e_d B_l 
     = \e_d^2 (A_l^{-1} + B_l^{-1})^{-1}. 
\ee
These small eigenvalues correspond to masses squared of 
charged leptons($e$, $\mu$, $\tau$). 
Unitary transformations which diagonalize $(A_l + B_l)$ and 
$\e_d^2 (A_l^{-1} + B_l^{-1})^{-1}$ are written as ${\cal W}_l$ 
and ${\cal V}_l$, namely 
\bea
  {\cal W}_l^{-1} (A_l + B_l) {\cal W}_l 
                    & = & (\Lambda _l^{(0)})^2, \\
  \e_d^2 \,{\cal V}_l^{-1} (A_l^{-1} + B_l^{-1})^{-1} 
                 {\cal V}_l & = & \e_d^2 \, (\Lambda _l^{(2)})^2, 
\label{eqn:WlVl} 
\eea
where $\Lambda _l^{(0)}$ and $\Lambda _l^{(2)}$ are diagonal. 
Thus explicit forms of the unitary matrices $\widehat{\cal V}_l$ and 
$\widehat{\cal U}_l$ in Eq.(\ref{eqn:Ml}) are 
\begin{eqnarray} 
   \widehat{\cal V}_l & \simeq & \left( 
   \begin{array}{cc} 
      y_S H {\cal W}_l \,{\Lambda _l^{(0)}}^{-1}  
         &  -y_S^{-1} {H^{\dag}}^{-1} {\cal V}_l 
                     \,\Lambda _l^{(2)}  \\
      y_N M {\cal W}_l \,{\Lambda _l^{(0)}}^{-1}  
         &  y_N^{-1} {M^{\dag}}^{-1} {\cal V}_l 
                         \,\Lambda _l^{(2)} \\
   \end{array} 
                       \right), 
\label{eqn:hatvl}                              \\
   \widehat{\cal U}_l & \simeq & \left( 
   \begin{array}{cc} 
      {\cal W}_l   &  -\epsilon _d (A_l + B_l)^{-1} 
                                     B_l {\cal V}_l \\
      \epsilon _d B_l (A_l + B_l)^{-1} {\cal W}_l    &  
                                           {\cal V}_l 
   \end{array} 
                       \right) 
\label{eqn:hatul}
\end{eqnarray} 
in the $\e_d$ expansion. 
{}From Eq. (\ref{eqn:hatvl}), the mass eigenstates of 
light $SU(2)_L$-doublet charged leptons are given by 
\be
   | \tilde{L}^{-} \rangle = \Lambda _l^{(2)} {\cal V}_l^T 
           \left( - y_S^{-1} {H^*}^{-1} | H_d^{-} \rangle 
                 + y_N^{-1} {M^*}^{-1} | L^{-} \rangle \right). 
\label{eqn:L-}
\ee
Consequently, provided that the elements of $y_S^{-1} {H^*}^{-1}$ and 
$y_N^{-1} {M^*}^{-1}$ are comparable to each other, 
there occur large mixings between $H_d^-$ and $L^-$. 
In order to parametrize the relative magnitude of $L^-$- 
and $H_d^-$-components we introduce the notation 
\be
  r_l  = \frac{y_S}{y_N} \,x^{\xi}
       = \frac{\langle S_0 \rangle}{\langle N^c_0 \rangle} \,x^{\xi}
       \sim \frac{y_S H_{33}}{y_N M_{33}} . 
\label{eqn:rl}
\ee

We now proceed to calculate the eigenvalues of 
$\e_d^2 (A_l^{-1} + B_l^{-1})^{-1}$. 
Generally, when a $3 \times 3$ Hermite matrix $C$ 
has hierarchical pattern as shown in Eq.(\ref{eqn:Mu}), 
three eigenvalues of the matrix $C$ are approximately expressed as 
\be
  {\rm Tr}(C), \qquad \qquad 
  \frac{\sum_i \Delta (C)_{ii}}{{\rm Tr}(C)}, \qquad \qquad 
  \frac{\det C}{\sum_i \Delta (C)_{ii}}, 
\label{eqn:C}
\ee
where $\Delta (C)_{ij}$ represents the cofactor for the $(i,j)$ 
element of $C$. 
When applied to $(A_l^{-1} + B_l^{-1})$, 
hierarchies of the eigenvalues depend on 
the relative magnitude of the elements of $A_l^{-1}$ and $B_l^{-1}$, 
which are controlled by $r_l$ as 
\begin{equation}
(A_l^{-1}+B_l^{-1})_{ij} = 
      y_N^{-2} x^{-(\beta_i+\beta_j)} \sum_k \{
r_l^{-2}x^{-2\beta_k}\overline{h}^*_{ki}\overline{h}_{kj} + 
        x^{-2\alpha_k}\overline{m}^*_{ki}\overline{m}_{kj} \}\ ,
\end{equation}
where we denote $\overline{h}_{ij}=(H_0^{\dagger -1})_{ij}$ and  
$\overline{m}_{ij}=(M_0^{\dagger -1})_{ij}$. 
Let us consider the following four regions of the parameter $r_l$,  
provided that the phenomenological conditions $m_e \geq \O(\l^9 v_d)$ 
and $m_{\tau} < \O(v_d)$ are satisfied. 

\begin{description}
\item[Case (i)] $x^{\a_1 - \b_2} (=\l) \leq r_l < x^{\a_2 - \b_2} (=1)$ \\
{}From Eq.(\ref{eqn:C}) light charged lepton masses become 
\bea
    m_e      & \sim & r_l x^{2\b_1} v_d   = r_l \l^8 v_d, \nonumber \\
    m_{\mu}  & \sim & x^{\a_1 + \b_2} v_d = \l^5 v_d, \\
    m_{\tau} & \sim & r_l x^{\b_2} v_d    = r_l \l^2 v_d. \nonumber
\eea

\item[Case (ii)] $x^{\a_2 - \b_2} (=1) \leq r_l 
                             < x^{\a_1 - \b_1} (=\l^{-1})$ \\
Charged lepton masses become 
\bea
    m_e      & \sim & r_l x^{2\b_1} v_d   = r_l \l^8 v_d, \nonumber \\
    m_{\mu}  & \sim & x^{\a_1 + \b_2} v_d = \l^5 v_d, \\
    m_{\tau} & \sim & x^{\a_2} v_d        = \l^2 v_d. \nonumber
\eea

\item[Case (iii)] $x^{\a_1 - \b_1} (=\l^{-1}) \leq r_l < 
                          x^{\a_2 - \b_1} (=\l^{-2})$ \\
In this region we obtain 
\bea
    m_e      & \sim & x^{\a_1 + \b_1} v_d     = \l^7 v_d, \nonumber \\
    m_{\mu}  & \sim & r_l x^{\b_1 + \b_2} v_d = r_l \l^6 v_d, \\
    m_{\tau} & \sim & x^{\a_2} v_d            = \l^2 v_d. \nonumber
\eea

\item[Case (iv)] $x^{\a_2 - \b_1} (=\l^{-2}) \leq r_l 
                               < x^{- \b_1} (=\l^{-4})$ \\ 
In this region we have 
\bea
    m_e      & \sim & x^{\a_1 + \b_1} v_d = \l^7 v_d, \nonumber \\
    m_{\mu}  & \sim & x^{\a_2 + \b_2} v_d = \l^4 v_d, \\
    m_{\tau} & \sim & r_l x^{\b_1} v_d    = r_l \l^4 v_d. \nonumber
\eea
\end{description}
Thus mass hierarchy of charged leptons apparently changes depending 
on the parameter $r_l$.
Experimentally the hierarchical charged lepton masses are summarized as 
\begin{equation}
\frac{m_e}{m_\mu}\simeq \l^{3.5} \qquad \frac{m_\mu}{m_\tau}\simeq \l^2 \ .
\end{equation} 
Among the above solutions the case (ii) gives rather large hierarchy 
$m_\mu/m_\tau\simeq \l^3$ as given in Eq.(26). 
For other three cases we obtain reasonable hierarchies for 
$m_e/m_\mu$ and $m_\mu/m_\tau$ by adjusting $r_l$ as 
\begin{eqnarray}
\frac{m_e}{m_\mu}&\simeq& \l^{3.5} \qquad 
\frac{m_\mu}{m_\tau}\simeq \l^{2.5}\quad  
      {\rm for\  the\  case\  (i)} \quad (r_l\simeq \sqrt{\l}) \nonumber\\
\frac{m_e}{m_\mu}&\simeq& \l^{3} \qquad 
\frac{m_\mu}{m_\tau}\simeq \l^{2}\quad  
      {\rm for\  the\  cases\  (iii) {\rm \ and\ } (iv)} \quad (r_l\simeq \l^{-2}) \ .
\end{eqnarray}  
In the followings we investigate the neutral lepton sector to combine 
the charged lepton solutions with light neutrino ones. 

\vspace{1cm}

In the neutral lepton sector we have the $15 \times 15$ 
mass matrix 
\begin{equation} 
\begin{array}{r@{}l} 
   \vphantom{\bigg(}   &  \begin{array}{cccccc} 
          \quad \, H_u^0   &  \ \  H_d^0  &  \quad L^0  
                          &  \quad \  N^c   &  \quad  S  &
        \end{array}  \\ 
\widehat{M}_N = 
   \begin{array}{l} 
        H_u^0  \\  H_d^0  \\  L^0  \\  N^c  \\  S  \\
   \end{array} 
     & 
\left( 
  \begin{array}{ccccc} 
       0     &   y_S H   &   y_N M^T   
                           &      0     &  \rho _d M^T  \\
     y_S H   &     0     &      0      
                           &      0     &  \rho _u M^T  \\
     y_N M   &     0     &      0      
                           &  \rho _u M &       0       \\
       0     &     0     & \rho _u M^T 
                           &      N     &      T^T      \\
   \rho _d M & \rho _u M &      0      
                           &      T     &       S       \\
  \end{array} 
\right) 
\end{array} 
\label{eqn:Mns} 
\end{equation} 
in $M_S$ units, 
where $\rho _u = \langle H_{u0}\rangle/M_S = v_u/M_S$. 
The $6 \times 6$ submatrix 
\begin{equation}
   \widehat{M}_M = \left(
   \begin{array}{cc}
        N    &   T^T  \\
        T    &    S     
   \end{array}
   \right) 
\label{eqn:MM}
\end{equation}
represents the Majorana mass terms which come from 
the nonrenormalizable interactions 
$(\Phi_i {\bar \Phi})(\Phi_j {\bar \Phi})(\Phi_0 {\bar \Phi})^{l_{ij}}$ 
with non-negative integers $l_{ij}$. 
Since matter fields $N^c_i$ and $S_i$ reside in the multipltes 
$\psi({\bf {\bar 6},2})_i$ and $\phi({\bf 15,1})_i$, respectively, 
this matrix has hierarchical structure. 
If the magnitude of the Majorana mass terms is  large enough 
compared to the electroweak scale, 
due to seesaw mechanism we can obtain small neutrino masses. 
By recalling the above study in the charged lepton sector, 
it is easy to see that the unitary matrix 
$\widehat{\cal U}_N$ which diagonalizes $\widehat{M}_N$ 
can be approximately decomposed into three factors as 
\be
  \widehat{\cal U}_N = \widehat{\cal U}_N^{(0)} \,
                       \widehat{\cal U}_N^{(1)} \,
                       \widehat{\cal U}_N^{(2)}, 
\ee
where the matrix $\widehat{\cal U}_N^{(0)}$ is essentially the same 
as the diagonalization matrix for light charged leptons and 
\be
 \widehat{\cal U}_N^{(1)} \simeq 
   \left( 
   \begin{array}{cc}
      I_{9 \times 9}   &         0             \\
             0         &  \widehat{\cal U}_M   
   \end{array}
   \right), \qquad 
 \widehat{\cal U}_N^{(2)} \simeq 
   \left( 
   \begin{array}{ccc}
       I_{6 \times 6}  &     0     &        0        \\
             0         & {\cal V}  &        0        \\
             0         &     0     &  I_{6 \times 6}   
   \end{array}
   \right). 
\ee
The matrix $\widehat{\cal U}_N^{(1)}$ means the diagonalization 
matrix for the Majorana mass matrix (\ref{eqn:MM}). 
The matrix $\widehat{\cal U}_N^{(2)}$ represents a diagonalization 
matrix for light neutrinos. 
It turns out that the light neutrino mass eigenstates are 
\be
   | \tilde{L}^0 \rangle = {\cal V}^T \Lambda _l^{(2)} {\cal V}_l^T 
           \left( - y_S^{-1} {H^*}^{-1} | H_d^0 \rangle 
                 + y_N^{-1} {M^*}^{-1} | L^0 \rangle \right). 
\label{eqn:L0}
\ee
Comparing these eigenstates $\tilde{L}^0$ with those of 
light charged leptons $\tilde{L}^-$ given 
by Eq. (\ref{eqn:L-}), 
we find that ${\cal V}$ is nothing but MNS matrix. 
The matrix ${\cal V}$ represents an additional transformation for 
neutrinos on the mass-diagonal basis for light charged leptons. 
${\cal V}$ is determined as the diagonalization matrix for 
the neutrino mass matrix $M_{\nu}$ defined by 
\be
  M_{\nu} = M_S \, \e_u^2 \, \left( \Lambda _l^{(2)} {\cal V}_l^{-1} 
           R^{-1} {\cal V}_l^* \Lambda _l^{(2)} \right), 
\label{eqn:Mnu}
\ee
where $\e_u = v_u/\langle N^c_0 \rangle$ and $R$ is the induced 
Majorana mass matrix stemming from Eq.(\ref{eqn:MM}). 
Note that the matrix $R$ has the hierarchical structure 
given by 
\be
R =  y_R \left(
  \begin{array}{ccc}
 r_{11}x^{2\b_1}     & r_{12}x^{\b_1+\b_2} & r_{13}x^{\b_1} \\
 r_{21}x^{\b_1+\b_2} & r_{22}x^{2\b_2}     & r_{23}x^{\b_2} \\
 r_{31}x^{\b_1}      & r_{32}x^{\b_2}      & r_{33} 
  \end{array}
  \right)
\label{eqn:R}
\ee
with symmetric $\O(1)$ numbers $r_{ij}$ and $M_S \,y_R$ represents 
the Majorana mass scale. 
As seen from Eq.(\ref{eqn:Mnu}) Dirac mass hierarchies given by 
$\Lambda _l^{(2)}$ cancel out in part or in large part 
due to seesaw mechanism. 
Since Dirac mass hierarchies depend on the parameter $r_l$, 
the neutrino mass matrix also depends on $r_l$. 
Thus we consider the following four cases separately. 

\begin{description}
\item[Case (i)] $x^{\a_1 - \b_2} (=\l) \leq r_l < x^{\a_2 - \b_2} (=1)$ \\
In this case the neutrino mass matrix becomes 
\be
   M_{\nu} =  \frac {v_u^2}{M_S \, y_R} \times 
      \left( 
      \begin{array}{ccc}
          O(r_l^2 x^{2\b_1}) & O(r_l x^{\a_1 + \b_1}) 
                               & O(r_l^2 x^{\b_1 + \b_2 }) \\
          O(r_l x^{\a_1 + \b_1 })  & O(x^{2\a_1})       
                               & O(r_l x^{\a_1 + \b_2})       \\
          O(r_l^2 x^{\b_1 + \b_2})   & O(r_l x^{\a_1 + \b_2})   
                               & O(r_l^2 x^{2\b_2})   
      \end{array}
      \right). 
\label{eqn:Mnu1}
\ee
This leads to the mixing angles in MNS matrix 
\bea
  \tan \theta_{12} & \sim & r_l x^{\b_1 - \a_1} 
                                      = r_l \l,  \nonumber \\
  \tan \theta_{23} & \sim & r_l^{-1} x^{\a_1 - \b_2} = r_l^{-1} \l,  \\
  \tan \theta_{13} & \sim & x^{\b_1 - \b_2} =  \l^2, \nonumber
\label{eqn:theta1}
\eea
where $\theta_{ij}$'s are defined as 
\be
   U_{MNS} = \left( 
       \begin{array}{ccc}
    c_{12} c_{13}       &       s_{12} c_{13}    &      s_{13}     \\
   -s_{12} c_{23} - c_{12} s_{23} s_{13}  
         & c_{12} c_{23} - s_{12} s_{23} s_{13}  &   s_{23} c_{13}  \\
    s_{12} s_{23} - c_{12} c_{23} s_{13} 
         & -c_{12} s_{23} - s_{12} c_{23} s_{13} &  c_{23} c_{13} 
       \end{array}
       \right). 
\ee
The ratios of neutrino masses are given by 
\be
  m_1 : m_2 : m_3 \sim 
   r_l^2 x^{2\b_1} : x^{2\a_1} : r_l^2 x^{2\b_2} 
   = r_l^2 \l^8 : \l^6 : r_l^2 \l^4. 
\ee

\item[Case (ii)] $x^{\a_2 - \b_2} (=1) \leq r_l 
                           < x^{\a_1 - \b_1} (=\l^{-1})$ \\
In this case the neutrino mass matrix becomes 
\be
   M_{\nu} =  \frac {v_u^2}{M_S \, y_R} \times 
      \left( 
      \begin{array}{ccc}
          O(r_l^2 x^{2\b_1}) & O(r_l x^{\a_1 + \b_1}) 
                               & O(r_l x^{\a_2 + \b_1 }) \\
          O(r_l x^{\a_1 + \b_1 })  & O(x^{2\a_1})       
                               & O(x^{\a_1 + \a_2})       \\
          O(r_l x^{\a_2 + \b_1})   & O(x^{\a_1 + \a_2})   
                               & O(x^{2\a_2})   
      \end{array}
      \right). 
\label{eqn:Mnu2}
\ee
This leads to the mixing angles in MNS matrix 
\bea
  \tan \theta_{12} & \sim & r_l x^{\b_1 - \a_1} 
                                      = r_l \l,  \nonumber \\
  \tan \theta_{23} & \sim &     x^{\a_1 - \a_2} =     \l,  \\
  \tan \theta_{13} & \sim & r_l x^{\b_1 - \a_2} = r_l \l^2. \nonumber
\label{eqn:theta2}
\eea
The ratios of neutrino masses are given by 
\be
  m_1 : m_2 : m_3 \sim 
   r_l^2 x^{2\b_1} : x^{2\a_1} : x^{2\a_2} 
   = r_l^2 \l^8 : \l^6 : \l^4. 
\ee

\item[Case (iii)] $x^{\a_1 - \b_1} (=\l^{-1}) \leq r_l < 
                            x^{\a_2 - \b_1} (=\l^{-2})$ \\
In this case the neutrino mass matrix becomes 
\be
   M_{\nu} =  \frac {v_u^2}{M_S \, y_R} \times 
      \left( 
      \begin{array}{ccc}
          O(x^{2\a_1}) & O(r_l x^{\a_1 + \b_1}) 
                               & O(x^{\a_1 + \a_2})     \\
          O(r_l x^{\a_1 + \b_1})  & O(r_l^2 x^{2\b_1})       
                               & O(r_l x^{\a_2 + \b_1}) \\
          O(x^{\a_1 + \a_2})   & O(r_l x^{\a_2 + \b_1})   
                               & O(x^{2\a_2})   
      \end{array}
      \right). 
\label{eqn:Mnu3}
\ee
We obtain the mixing angles 
\bea
  \tan \theta_{12} & \sim & r_l^{-1} x^{\a_1 - \b_1} 
                                   = (r_l \l)^{-1}, \nonumber \\
  \tan \theta_{23} & \sim & r_l x^{\b_1 - \a_2} = r_l \l^2, \\
  \tan \theta_{13} & \sim & x^{\a_1 - \a_2} = \l. \nonumber
\label{eqn:theta3}
\eea
The ratios of neutrino masses are 
\be
  m_1 : m_2 : m_3 \sim 
    x^{2\a_1} : r_l^2 x^{2\b_1} : x^{2\a_2} 
   = \l^6 : r_l^2 \l^8 : \l^4. 
\ee

\item[Case (iv)] $x^{\a_2 - \b_1} (=\l^{-2}) \leq r_l 
                              < x^{- \b_1} (=\l^{-4})$ \\
In this case the neutrino mass matrix becomes 
\be
   M_{\nu} =  \frac {v_u^2}{M_S \, y_R} \times 
      \left( 
      \begin{array}{ccc}
          O(x^{2\a_1}) & O(x^{\a_1 + \a_2}) 
                               & O(r_l x^{\a_1 + \b_1 }) \\
          O(x^{\a_1 + \a_2})  & O(x^{2\a_2})       
                               & O(r_l x^{\a_2 + \b_1})  \\
          O(r_l x^{\a_1 + \b_1})   & O(r_l x^{\a_2 + \b_1})   
                               & O(r_l^2 x^{2\b_1})   
      \end{array}
      \right). 
\label{eqn:Mnu4}
\ee
We have the mixing angles 
\bea
  \tan \theta_{12} & \sim & x^{\a_1 - \a_2} = \l, \nonumber \\
  \tan \theta_{23} & \sim & r_l^{-1} x^{\a_2 - \b_1} = (r_l \l^2)^{-1}, \\
  \tan \theta_{13} & \sim & r_l^{-1} x^{\a_1 - \b_1} = (r_l \l)^{-1}. 
                                       \nonumber 
\label{eqn:theta4}
\eea
The ratios of neutrino masses are 
\be
  m_1 : m_2 : m_3 \sim 
    x^{2\a_1} : x^{2\a_2} : r_l^2 x^{2\b_1} 
   = \l^6 : \l^4 : r_l^2 \l^8. 
\ee

\end{description}

As mentioned above, the characteristic pattern of neutrino 
masses and mixing angles varies significantly depending on 
the parameter $r_l$. 
As seen from Eq.(\ref{eqn:rl}), the choice of $r_l$ corresponds to 
the adjustment of $\xi (= 2b_3 + a_0)$. 
A large mixing angle solution in which both $\tan \theta_{12}$ and 
$\tan \theta_{23}$ are $\O(1)$ can be realized by taking 
$r_l \sim \l^{-1.5}$ in the case (iii). 
Otherwise, at least one of $\tan \theta_{12}$ and $\tan \theta_{23}$ 
becomes small. 
In Ref.\cite{Matsu2} we adopted the value of $r_l = 1$ in the case (i) or (ii) 
and then we obtained slightly small mixing angles, i.e., 
$\tan \theta_{12}, \ \tan \theta_{23} = \O(\l)$. 
Instead we choose 
\be
   r_l = \l^{-1.5} 
\ee
in the case (iii). 
This choice of $r_l$ together with $e_{33} =0$ exhibits 
constraints on the flavor charge assignment for matter fields. 
It should be emphasized that in the case (iii) Dirac mass 
hierarchies cancel out with $R^{-1}$ in large part 
due to seesaw mechanism. 
Thus we have the lepton mass hierarchies 
\bea 
    m_e       \sim \l^7 v_d,     \phantom{MMMM} 
    m_{\mu}   \sim \l^{4.5} v_d, \phantom{MMMM} 
    m_{\tau}  \sim \l^2 v_d,                        \\
    m_1       \sim \l^6 \frac{v_u^2}{M_S \, y_R}, \qquad 
    m_2       \sim \l^5 \frac{v_u^2}{M_S \, y_R}, \qquad 
    m_3       \sim \l^4 \frac{v_u^2}{M_S \, y_R}, 
\eea
which lead to the numerical values 
\bea
    m_e       \sim 2\,{\rm MeV},   \phantom{MMM} 
    m_{\mu}   \sim 100\,{\rm MeV}, \phantom{MMM} 
    m_{\tau}  \sim 5\,{\rm GeV},                  \\
    m_1       \sim 0.002\,{\rm eV}, \phantom{MMM} 
    m_2       \sim 0.01\,{\rm eV},  \phantom{MMM} 
    m_3       \sim 0.05\,{\rm eV}, 
\eea
provided that $v_u,\ v_d \sim 100{\rm GeV}$ and 
$M_S y_R = 5 \times 10^{11}{\rm GeV}$. 
Further, the mixing angles become 
\bea
  \tan \theta_{12} \sim \sqrt{\l}, \qquad 
  \tan \theta_{23} \sim \sqrt{\l}, \qquad 
  \tan \theta_{13} \sim \l. 
\label{eqn:theta}
\eea
If $\tan \theta = \sqrt{\l} = 0.47$, 
then we obtain $\sin ^22\theta = 0.59$. 
Since Eq.(\ref{eqn:theta}) is an order of magnitude relationship, 
the above results are consistent with the recent data 
on atmospheric and solar neutrinos. 
As for neutrino mass differences, we have the ratio 
\be
  \frac{\Delta m_{12}^2}{\Delta m_{23}^2} 
  \sim \frac{m_2^2}{m_3^2} 
  \sim \l^2 \sim \frac{1}{20}. 
\ee
This is also consistent with the data. 
In this model flavor mixings come from mainly the neutrino 
mass matrix because the charged lepton one is hierarchical 
and ${\cal V}_l \sim 1$, which is similar to the 
diagonalization matrix for quark mass matrices\cite{CKM}. 
One of the characteristic features of the mixing matrix 
obtained here is to give a relatively large value for $U_{e3}$, 
which is $\sim \l$ near to the experimental bound of 
CHOOZ($\leq 0.16$)\cite{CHOOZ}, 
in contrast to $V_{CKM,td}(\sim \l^3)$ or 
$V_{CKM,ub}(\sim \l^4)$ in the quark mixing matrix. 
The bi-maximal mixing solution usually gives the tiny magnitude for $U_{e3}$\cite{Bimax}.
The measurement of $U_{e3}$ will give an important clue to 
distinguish the various models.  
The reactor experiment, for example, KamLAND\cite{KamLAND} 
will cover the LMA-MSW region and possibly give the restriction 
on $U_{e3}$ through $\nu_e$ disappearance experiment as 
\bea
  \sum_{x=\mu,\tau}P(\nu_e\ra \nu_x) & = & 4|U_{e3}|^2 
         (1-|U_{e3}|^2) \sin^2 \frac{\Delta m^2_{atm}L}{4E}, 
                                                 \nonumber \\
  \sin^2 2\theta_{reac} & = & 4|U_{e3}|^2 (1-|U_{e3}|^2) 
         \simeq 0.09 \quad ({\rm for}\ \ \ |U_{e3}|\simeq 0.15). 
\eea

As another choice of the parameter $r_l$ it is interesting 
for us to take $r_l \sim \sqrt{\l}$ in the case (i). 
In this choice lepton masses become 
\bea 
    m_e       \sim \l^{8.5} v_d,     \phantom{MMMM} 
    m_{\mu}   \sim \l^5 v_d,         \phantom{MMMM} 
    m_{\tau}  \sim \l^{2.5} v_d,                        \\
    m_1       \sim \l^9 \frac{v_u^2}{M_S \, y_R}, \qquad 
    m_2       \sim \l^6 \frac{v_u^2}{M_S \, y_R}, \qquad 
    m_3       \sim \l^5 \frac{v_u^2}{M_S \, y_R}. 
\eea
Numerically by taking $M_S y_R = 2 \times 10^{12}{\rm GeV}$ we obtain 
\bea
    m_e       \sim 0.3\,{\rm MeV},   \phantom{MMM} 
    m_{\mu}   \sim 50\,{\rm MeV},    \phantom{MMM} 
    m_{\tau}  \sim 2\,{\rm GeV},                  \\
    m_1       \sim 0.0001\,{\rm eV}, \phantom{MMM} 
    m_2       \sim 0.01\,{\rm eV},   \phantom{MMM} 
    m_3       \sim 0.05\,{\rm eV}. 
\eea
Further, we have the mixing angles 
\bea
  \tan \theta_{12} \sim \l^{1.5}, \qquad 
  \tan \theta_{23} \sim \sqrt{\l}, \qquad 
  \tan \theta_{13} \sim \l^2. 
\label{eqn:thetaB}
\eea
The magnitudes of $\tan \theta_{23}$ and 
$\Delta m_{12}^2/\Delta m_{23}^2$  are the same as in 
the above solution with $r_l = \l^{-1.5}$. 
While the magnitude of $\tan \theta _{12}$ is near 
the small mixing angle MSW region for $\nu_e$-$\nu_{\mu}$ 
oscillation.

In conclusion, the present model can yield a large mixing angle 
solution with $\tan \theta_{12}$, $\tan \theta_{23} = \O(\sqrt{\l})$ 
together with lepton mass hierarchies. 
Dirac mass hierarchies in the neutrino sector cancel out with the heavy Majorana sector  
in large part due to seesaw mechanism. 
Hierarchical pattern of charged leptons is also explained. 
In the string inspired models the massless sector contains 
extra particles beyond the minimal supersymmetric standard model. 
In the course of the gauge symmetry breakings many particles 
become massive or are absorbed by gauge fields via Higgs mechanism 
at intermediate energy scales. 
Therefore, after integrating out these heavy modes we derive 
the low-energy effective theory 
in which large extra-particle mixings cause an apparent change 
of the Yukawa hierarchies for leptons and down-type quarks. 
In the neutrino sector seesaw mechanism is also incorporated. 
This is the reason why nontrivial patterns appear in 
fermion mixing angles. 
Finally, we comment on the flavor symmetry. 
In the present model we need an appropriate discrete flavor 
symmetry and also the adjustment of the flavor charge assignment 
for matter fields. 
In our study we assign appropriate flavor charges to matter fields 
by hand so as to obtain an interesting solution. 
In the framework of string theory it is important for us to 
explore the selection rule including the flavor symmetry.

\section*{Acknowledgements}
Authors would like to thank valuable discussions at 
the research meeting on neutrino physics held at Fuji-yoshida 
on 21st to 24th, August, 2000. 
The authors are supported in part by a Grant-in-Aid 
for scientific Research from Ministry of Education, Science, 
Sports and Culture, Japan (No. 12047226).



\begin{thebibliography}{99}

\bibitem{Atmos}
Super-Kamiokande Collab., Y. Fukuda et. al., \PRL{81}{1562}{1998}; 
\PL{B436}{33}{1998}; \PRL{82}{2644}{1999} \\
H. Sobel, talk given at the 19th International Conference on 
Neutrino Physics and Astrophysics, Sudbury, Canada, 
June 16-21, 2000. 

\bibitem{Solar}
Y. Suzuki, talk given at the 19th International Conference on 
Neutrino Physics and Astrophysics, Sudbury, Canada, 
June 16-21, 2000. 

\bibitem{Seesaw}
T. Yanagida, in {\it Proceedings of the Workshop 
    on Unified Theory and Baryon Number in the 
    Universe}, ed. O. Sawada and A. Sugamoto 
    (KEK, report 79-18, 1979), p. 95. \\
M. Gell-Mann, P. Ramond and S. Slansky, 
    in {\it Supergravity}, 
    ed. P. van Nieuwenhuizen and D. Z. Freedman  
    (North-Holland, Amsterdam, 1979), p. 315. \\
R. Mohapatra and S. Senjanovi\'c, Phys. Rev. Lett. 
         {\bf 44} (1980), 912. 

\bibitem{CKM}
M. Matsuda and T. Matsuoka, \PL{B487}{104}{2000}.

\bibitem{Matsu1}
N. Haba, C. Hattori, M. Matsuda and T. Matsuoka, \PTP{96}{1249}{1996}.

\bibitem{Matsu2}
N. Haba and T. Matsuoka, \PTP{99}{831}{1998}.

\bibitem{Matsu3}
T. Matsuoka, \PTP{100}{107}{1998}.

\bibitem{Matsu4}
N. Haba, C. Hattori, M. Matsuda, T. Matsuoka and D. Mochinaga, 
         \PTP{94}{233}{1995}.

\bibitem{Pati}
J. C. Pati and A. Salam, \PR{D10}{275}{1974}.

\bibitem{Scale}
N. Haba, C. Hattori, M. Matsuda, T. Matsuoka and D. Mochinaga, 
         \PL{B337}{63}{1994}; \PTP{92}{153}{1994}.

\bibitem{F-N}
C. Froggatt and H. B. Nielsen, \NP{B147}{277}{1979}.

\bibitem{CHOOZ}
CHOOZ Collab., M. Apollonio {\it et. al.}, \PL{B466}{415}{1999}.

\bibitem{Bimax}
V. Barger, S. Pakvasa, T. J. Weiler and K. Whisnant, \PL{B437}{107}{1998}.\\
A. J. Baltz, A. S. Goldhaber and M. Goldhaber, \PRL{81}{5730}{1998}.\\
M. Jezabek and Y. Sumino, \PL{B440}{327}{1998}.\\
R. N. Mohapatra and S. Nussinov, \PL{B441}{299}{1998}.\\
C. Jarlskog, M. Matsuda, S. Skadhauge and M. Tanimoto, \PL{B449}{240}{1999}.\\
Y. Nomura and T. Yanagida, \PR{D59}{017303}{1999}. 
\bibitem{KamLAND}
A. Piepke, talk given at the 19th International Conference on 
Neutrino Physics and Astrophysics, Sudbury, Canada, 
June 16-21, 2000. 

\end{thebibliography}
\end{document}